\documentclass[aps,showpacs,showkeys,amsmath,amssymb,pra,reprint,superscriptaddress,lengthcheck]{revtex4-1}

\usepackage{graphicx}
\usepackage{dcolumn}
\usepackage{bm}
\usepackage{color}
\usepackage{amsmath}
\usepackage{amssymb}
\usepackage{gensymb}
\usepackage{slashed}
\usepackage{hyperref}
\hypersetup{colorlinks=true,
            linkcolor=blue,
            urlcolor=blue,
            citecolor=blue}

\definecolor{aogreen}{rgb}{0.0, 0.5, 0.0}
\definecolor{RiaanGreen}{RGB}{45,119,0}

\definecolor{revisedcolor}{RGB}{0,100,20}

\definecolor{Bcolor}{RGB}{10,200,10}
\definecolor{Jcolor}{RGB}{20,20,200}
\definecolor{Ccolor}{RGB}{200,20,20}
\definecolor{Qcolor}{RGB}{50,200,200}

\begin{document}

@book{Budker:2013,
  title={Optical Magnetometry},
  author={edited by D. Budker and D. F. Jackson Kimball},
  year={2013},
  publisher={Cambridge University Press, Cambridge, England}
}

@article{Budker/N:2007,
  title={Optical magnetometry},
  author={Budker, Dmitry and Romalis, Michael},
  journal={Nat. Phys.},
  volume={3},
  pages={227--234},
  year={2007},
  doi={10.1038/nphys566}
}

@article{Johnson/PMB:2013,
doi = {10.1088/0031-9155/58/17/6065},
url = {https://dx.doi.org/10.1088/0031-9155/58/17/6065},
year = {2013},
month = {aug},
publisher = {IOP Publishing},
volume = {58},
number = {17},
pages = {6065},
author = {Cort N Johnson and P D D Schwindt and M Weisend},
title = {Multi-sensor magnetoencephalography with atomic magnetometers},
journal = { Phys. Med. Biol.},
}

@article{Vasilakis/PRL:2009,
  title = {Limits on New Long Range Nuclear Spin-Dependent Forces Set with a $\mathbf{K}\mathrm{\text{\ensuremath{-}}}^{3}\mathrm{He}$ Comagnetometer},
  author = {Vasilakis, G. and Brown, J. M. and Kornack, T. W. and Romalis, M. V.},
  journal = {Phys. Rev. Lett.},
  volume = {103},
  issue = {26},
  pages = {261801},
  numpages = {4},
  year = {2009},
  month = {Dec},
  publisher = {American Physical Society},
  doi = {10.1103/PhysRevLett.103.261801},
  url = {https://link.aps.org/doi/10.1103/PhysRevLett.103.261801}
}

@article{Dang/APL:2010,
    author = {Dang, H. B. and Maloof, A. C. and Romalis, M. V.},
    title = "{Ultrahigh sensitivity magnetic field and magnetization measurements with an atomic magnetometer}",
    journal = { Appl. Phys. Lett.},
    volume = {97},
    number = {15},
    pages = {151110},
    year = {2010},
    month = {10},
    issn = {0003-6951},
    doi = {10.1063/1.3491215},
    url = {https://doi.org/10.1063/1.3491215}
}

@article{Shah/NP:2007,
  title={Subpicotesla atomic magnetometry with a microfabricated vapour cell},
  author={Shah, V. and Knappe, S. and Schwindt, P. D. D. and Kitching, J.},
  journal={Nat. Photonics},
  volume={1},
  pages={649},
  year={2007},
  doi={10.1038/nphoton.2007.201}
}

@article{Rubinsztein-Dunlop/JO:2017,
doi = {10.1088/2040-8978/19/1/013001},
url = {https://dx.doi.org/10.1088/2040-8978/19/1/013001},
year = {2017},
month = {nov},
publisher = {IOP Publishing},
volume = {19},
number = {1},
pages = {013001},
author = {Halina Rubinsztein-Dunlop and Andrew Forbes and M V Berry and M R Dennis and David L Andrews and Masud Mansuripur and Cornelia Denz and Christina Alpmann and Peter Banzer and Thomas Bauer and Ebrahim Karimi and Lorenzo Marrucci and Miles Padgett and Monika Ritsch-Marte and Natalia M Litchinitser and Nicholas P Bigelow and C Rosales-Guzmán and A Belmonte and J P Torres and Tyler W Neely and Mark Baker and Reuven Gordon and Alexander B Stilgoe and Jacquiline Romero and Andrew G White and Robert Fickler and Alan E Willner and Guodong Xie and Benjamin McMorran and Andrew M Weiner},
title = {Roadmap on structured light},
journal = {J. Opt.},
}

@book{Gbur:2017,
  title={Singular optics},
  author={Gbur, Gregory J},
  year={2017},
  publisher={CRC press}
}

@article{Castellucci/PRL:2021,
  title = {Atomic Compass: Detecting 3D Magnetic Field Alignment with Vector Vortex Light},
  author = {Castellucci, Francesco and Clark, Thomas W. and Selyem, Adam and Wang, Jinwen and Franke-Arnold, Sonja},
  journal = {Phys. Rev. Lett.},
  volume = {127},
  issue = {23},
  pages = {233202},
  numpages = {6},
  year = {2021},
  month = {Nov},
  publisher = {American Physical Society},
  doi = {10.1103/PhysRevLett.127.233202},
  url = {https://link.aps.org/doi/10.1103/PhysRevLett.127.233202}
}

@article{Qiu/PR:2021,
author = {Shuwei Qiu and Jinwen Wang and Francesco Castellucci and Mingtao Cao and Shougang Zhang and Thomas W. Clark and Sonja Franke-Arnold and Hong Gao and Fuli Li},
journal = {Photon. Res.},
keywords = {CCD cameras; Cylindrical vector beams; Nitrogen vacancy centers; Optical fields; Tunable diode lasers; Vector beams},
number = {12},
pages = {2325--2331},
publisher = {Optica Publishing Group},
title = {Visualization of magnetic fields with cylindrical vector beams in a warm atomic vapor},
volume = {9},
month = {Dec},
year = {2021},
url = {https://opg.optica.org/prj/abstract.cfm?URI=prj-9-12-2325},
doi = {10.1364/PRJ.418522},
}

@article{Savukov/PRL:2005,
  title = {Tunable Atomic Magnetometer for Detection of Radio-Frequency Magnetic Fields},
  author = {Savukov, I. M. and Seltzer, S. J. and Romalis, M. V. and Sauer, K. L.},
  journal = {Phys. Rev. Lett.},
  volume = {95},
  issue = {6},
  pages = {063004},
  numpages = {4},
  year = {2005},
  month = {Aug},
  publisher = {American Physical Society},
  doi = {10.1103/PhysRevLett.95.063004},
  url = {https://link.aps.org/doi/10.1103/PhysRevLett.95.063004}
}

@article{Ledbetter/PRA:2007,
  title = {Detection of radio-frequency magnetic fields using nonlinear magneto-optical rotation},
  author = {Ledbetter, M. P. and Acosta, V. M. and Rochester, S. M. and Budker, D. and Pustelny, S. and Yashchuk, V. V.},
  journal = {Phys. Rev. A},
  volume = {75},
  issue = {2},
  pages = {023405},
  numpages = {6},
  year = {2007},
  month = {Feb},
  publisher = {American Physical Society},
  doi = {10.1103/PhysRevA.75.023405},
  url = {https://link.aps.org/doi/10.1103/PhysRevA.75.023405}
}

@book{Bransden:2003,
  title={Physics of Atoms and Molecules},
  author={Bransden, Brian Harold and Joachain, Charles Jean},
  year={2003},
  publisher={Prentice Hall, Harlow, England}
}

@book{Blum:2012,
  title={Density Matrix Theory and Applications},
  author={Blum, Karl},
  year={2012},
  publisher={Springer, Berlin}
}

@book{Auzinsh:2010,
  title={Optically Polarized Atoms: Understanding Light-Atom Interactions},
  author={Auzinsh, Marcis and Budker, Dmitry and Rochester, Simon M},
  year={2010},
  publisher={Oxford University, Oxford}
}

@article{Wense:2020,
  title={The theory of direct laser excitation of nuclear transitions},
  author={von der Wense, Lars and Bilous, Pavlo V and Seiferle, Benedict and Stellmer, Simon and Weitenberg, Johannes and Thirolf, Peter G and P{\'a}lffy, Adriana and Kazakov, Georgy},
  journal={Eur. Phys. J. A},
  volume={56},
  pages={176},
  year={2020},
  publisher={Springer},
  url = {https://doi.org/10.1140/epja/s10050-020-00177-x}
}

@article{Tremblay/PRA:1990,
  title = {Optical pumping with two finite linewidth lasers},
  author = {Tremblay, P. and Jacques, C.},
  journal = {Phys. Rev. A},
  volume = {41},
  issue = {9},
  pages = {4989--4999},
  numpages = {0},
  year = {1990},
  month = {May},
  publisher = {American Physical Society},
  doi = {10.1103/PhysRevA.41.4989},
  url = {https://link.aps.org/doi/10.1103/PhysRevA.41.4989}
}

@article{Schmidt/PRA:2024,
  title = {Atomic photoexcitation as a tool for probing purity of twisted light modes},
  author = {Schmidt, R. P. and Ramakrishna, S. and Peshkov, A. A. and Huntemann, N. and Peik, E. and Fritzsche, S. and Surzhykov, A.},
  journal = {Phys. Rev. A},
  volume = {109},
  issue = {3},
  pages = {033103},
  numpages = {11},
  year = {2024},
  month = {Mar},
  publisher = {American Physical Society},
  doi = {10.1103/PhysRevA.109.033103},
  url = {https://link.aps.org/doi/10.1103/PhysRevA.109.033103}
}

@article{Matula/JPB:2013,
doi = {10.1088/0953-4075/46/20/205002},
url = {https://dx.doi.org/10.1088/0953-4075/46/20/205002},
year = {2013},
month = {oct},
publisher = {IOP Publishing},
volume = {46},
number = {20},
pages = {205002},
author = {O Matula and A G Hayrapetyan and V G Serbo and A Surzhykov and S Fritzsche},
title = {Atomic ionization of hydrogen-like ions by twisted photons: angular distribution of emitted electrons},
journal = {J. Phys. B},
}

@article{Knyazev/PU:2018,
doi = {10.3367/UFNe.2018.02.038306},
url = {https://dx.doi.org/10.3367/UFNe.2018.02.038306},
year = {2018},
month = {may},
publisher = {Uspekhi Fizicheskikh Nauk, Russian Academy of Sciences and IOP Publishing},
volume = {61},
pages = {449},
author = {B A Knyazev and V G Serbo},
title = {Beams of photons with nonzero projections of orbital angular momenta: new results},
journal = {Phys.-Usp.},
}

@article{Schulz/PRA:2020,
  title = {Generalized excitation of atomic multipole transitions by twisted light modes},
  author = {Schulz, S. A.-L. and Peshkov, A. A. and M\"uller, R. A. and Lange, R. and Huntemann, N. and Tamm, Chr. and Peik, E. and Surzhykov, A.},
  journal = {Phys. Rev. A},
  volume = {102},
  issue = {1},
  pages = {012812},
  numpages = {10},
  year = {2020},
  month = {Jul},
  publisher = {American Physical Society},
  doi = {10.1103/PhysRevA.102.012812},
  url = {https://link.aps.org/doi/10.1103/PhysRevA.102.012812}
}

@book{Johnson:2007,
  title={Atomic Structure Theory},
  author={Johnson, Walter R},
  year={2007},
  publisher={Springer, New York}
}

@article{Solyanik-Gorgone/JOSAB:2019,
author = {Maria Solyanik-Gorgone and Andrei Afanasev and Carl E. Carlson and Christian T. Schmiegelow and Ferdinand Schmidt-Kaler},
journal = {J. Opt. Soc. Am. B},
keywords = {Circular polarization; Laser beams; Light beams; Light matter interactions; Optical vortices; Orbital angular momentum multiplexing},
number = {3},
pages = {565--574},
publisher = {Optica Publishing Group},
title = {Excitation of E1-forbidden atomic transitions with electric, magnetic, or mixed multipolarity in light fields carrying orbital and spin angular momentum \[Invited\]},
volume = {36},
month = {Mar},
year = {2019},
url = {https://opg.optica.org/josab/abstract.cfm?URI=josab-36-3-565},
doi = {10.1364/JOSAB.36.000565}
}

@article{Peshkov/AdP:2023,
author = {Peshkov, Anton A. and Jordan, Elena and Kromrey, Markus and Mehta, Karan K. and Mehlstäubler, Tanja E. and Surzhykov, Andrey},
title = {Excitation of Forbidden Electronic Transitions in Atoms by Hermite–Gaussian Modes},
journal = {Ann. Phys.},
volume = {535},
number = {9},
pages = {2300204},
doi = {https://doi.org/10.1002/andp.202300204},
url = {https://onlinelibrary.wiley.com/doi/abs/10.1002/andp.202300204},
year = {2023}
}

@book{Rose:1957,
  title={Elementary Theory of Angular Momentum},
  author={Rose, Morris Edgar},
  year={1957},
  publisher={John Wiley \& Sons, New York}
}

@article{Fritzsche/CPC:2019,
  title={A fresh computational approach to atomic structures, processes and cascades},
  author={Stephan Fritzsche},
  journal={Comput. Phys. Commun.},
  volume={240},
  number={1},
  pages={1-14},
  year={2019},
  doi = {10.1016/j.cpc.2019.01.012},
}

@article{Maldonado/OE:2024,
  title={Sensitivity of a vector atomic magnetometer based on electromagnetically induced transparency},
  author={Mario Gonzalez Maldonado and Owen Rollins and Alex Toyryla and James A. McKelvy and Andrey Matsko and Isaac Fan and Yang Li and Ying-Ju Wang and John Kitching and Irina Novikova and Eugeniy E. Mikhailov},
  journal={Opt. Express},
  volume={32},
  pages={25062-25073},
  year={2024},
  doi = {10.1364/OE.529276},
}

@article{Wang/AVSQS:2020,
  title={Vectorial light–matter interaction: Exploring spatially structured complex light fields},
  author={Jinwen Wang and Francesco Castellucci and Sonja {Franke-Arnold}},
  journal={AVS Quantum Sci.},
  volume={2},
  pages={031702},
  year={2020},
  doi = {10.1116/5.0016007},
}

@article{Wang/PRL:2024,
  title={Measuring the Optical Concurrence of Vector Beams with an Atomic-State Interferometer},
  author={Jinwen Wang and Sphinx J. Svensson and Thomas W. Clark and Yun Chen and Mustafa A. {Al Khafaji} and Hong Gao and Niclas Westerberg and Sonja {Franke-Arnold}},
  journal={Phys. Rev. Lett.},
  volume={132},
  pages={193803},
  year={2024},
  doi = {10.1103/PhysRevLett.132.193803},
}

@article{Volz/PHysScripta:1996,
  title={Precision lifetime measurements on alkali atoms and on helium by beam–gas–laser spectroscopy},
  author={U. Volz and H. Schmoranzer},
  journal={Physica Scripta},
  volume={T65},
  pages={48-56,},
  year={1996},
  doi = {10.1088/0031-8949/1996/T65/007},
}

\preprint{}
\title{
Interaction of vector light beams with atoms\\exposed to a time-dependent magnetic field
}

\author{S.~Ramakrishna}
\thanks{These two authors contributed equally to this work.}
\email[]{shreyas.ramakrishna@uni-jena.de}
\affiliation{Helmholtz-Institut Jena, D-07743 Jena, Germany}%
\affiliation{GSI Helmholtzzentrum f\"ur Schwerionenforschung GmbH, D-64291 Darmstadt, Germany}
\affiliation{Theoretisch-Physikalisches Institut, Friedrich-Schiller-Universit\"at Jena, D-07743 Jena, Germany}

\author{R.~P.~Schmidt}
\thanks{These two authors contributed equally to this work.}
\email[]{riaan.schmidt@ptb.de}
\affiliation{Physikalisch-Technische Bundesanstalt, D-38116 Braunschweig, Germany}
\affiliation{Institut für Mathematische Physik, Technische Universität Braunschweig, D-38106 Braunschweig, Germany}

\author{A.~A.~Peshkov}
\affiliation{Physikalisch-Technische Bundesanstalt, D-38116 Braunschweig, Germany}
\affiliation{Institut für Mathematische Physik, Technische Universität Braunschweig, D-38106 Braunschweig, Germany}

\author{S.~Franke-Arnold}
\affiliation{School of Physics and Astronomy, University of Glasgow, Glasgow G12 8QQ, United Kingdom}

\author{A.~Surzhykov}
\affiliation{Physikalisch-Technische Bundesanstalt, D-38116 Braunschweig, Germany}
\affiliation{Institut für Mathematische Physik, Technische Universität Braunschweig, D-38106 Braunschweig, Germany}
\affiliation{Laboratory for Emerging Nanometrology Braunschweig, D-38106 Braunschweig, Germany}

\author{S.~Fritzsche}
\affiliation{Helmholtz-Institut Jena, D-07743 Jena, Germany}%
\affiliation{GSI Helmholtzzentrum f\"ur Schwerionenforschung GmbH, D-64291 Darmstadt, Germany}
\affiliation{Theoretisch-Physikalisches Institut, Friedrich-Schiller-Universit\"at Jena, D-07743 Jena, Germany}

\date{\today}

\begin{abstract}
During recent years interest has been rising for applications of vector light beams towards magnetic field sensing. In particular, a series of experiments were performed to extract information about properties of static magnetic fields from absorption profiles of light passing through an atomic gas target. In the present work, we propose an extension to this method for oscillating magnetic fields. To investigate this scenario, we carried out theoretical analysis based on the time-dependent density matrix theory. We found that absorption profiles, even when averaged over typical observation times, are indeed sensitive to both strength and frequency of the time-dependent field, thus opening the prospect for a powerful diagnostic technique. To illustrate this sensitivity, we performed detailed calculations for the $5s \;\, {}^2S_{1/2}$ ($F=1$) $-$ $5p \;\, {}^2 P_{3/2}$ ($F=0$) transition in rubidium atoms, subject to a superposition of an oscillating (test) and a static (reference) magnetic field.

\end{abstract}

\newpage
\maketitle


\section{Introduction}\label{Sec.Intro}
In optical magnetometry, magnetic field properties are measured by observing changes in the optical properties of an atomic medium immersed in the field \cite{Budker:2013}. The great advantage of this technology compared to superconducting magnetic field sensors is that it offers high sensitivity without requiring cryogenic temperatures \cite{Budker/N:2007}. Optical magnetometers are finding applications in a wide variety of fields including medicine \cite{Johnson/PMB:2013}, fundamental physics \cite{Vasilakis/PRL:2009}, and geophysics \cite{Dang/APL:2010}. Significant progress has been made in miniaturizing these devices and improving their operating characteristics \cite{Shah/NP:2007}.

In most optical magnetometry experiments, the polarization across the light beam is approximately uniform. Meanwhile, recent advances in optics made it possible to generate light fields with space-varying polarization \cite{Rubinsztein-Dunlop/JO:2017}. The best known examples of such vector beams include radially and azimuthally polarized beams with an azimuthally varying linear polarization surrounding an optical vortex \cite{Gbur:2017}. These spatially varying light polarizations can excite locally varying magnetization profiles in atoms \cite{Wang/AVSQS:2020}. It has been demonstrated recently in cold \cite{Castellucci/PRL:2021} and warm \cite{Qiu/PR:2021,Maldonado/OE:2024} atomic vapors that such spatially varying light-matter interaction allows a simultaneous measurement of magnetic field components transverse and along the optical axis. These experiments determined static magnetic fields from absorption profiles of a vector beam after its passage through the rubidium vapor. We note that in this work as well as for the mechanisms proposed here the properties of the vector light change over length scales much larger than the extend of individual atoms.

Beyond static magnetic fields, the detection of time-dependent magnetic fields is important, especially in the radio-frequency domain. Detection of fields in the kilohertz to gigahertz frequency range finds many applications, from radio communication to detection of nuclear magnetic resonance (NMR) and nuclear quadrupole resonance (NQR) signals \cite{Savukov/PRL:2005}. An example of an atomic magnetometer for detection of radio-frequency magnetic fields is magnetometer \cite{Ledbetter/PRA:2007} based on a nonlinear magneto-optical rotation. In this paper, we explore the possibility of detecting oscillating (radio-frequency) magnetic fields based on another effect, namely the dependence of the absorption profile of a vector beam, propagating through an atomic vapor, on the strength and frequency of the magnetic field. In addition to the light field and the magnetic field to be measured, here we need an additional reference static magnetic field with well-known direction and strength. As in previous works \cite{Savukov/PRL:2005, Ledbetter/PRA:2007}, the static magnetic field is perpendicular to the oscillating one. While the static field in Refs.~\cite{Savukov/PRL:2005, Ledbetter/PRA:2007} was added to achieve the Zeeman resonance of atoms, it is here applied in order to trigger oscillations in the orientation of the resulting magnetic field relative to the light polarization.

\begin{figure*}[t]
	\centering
	\includegraphics[width=0.75\textwidth]{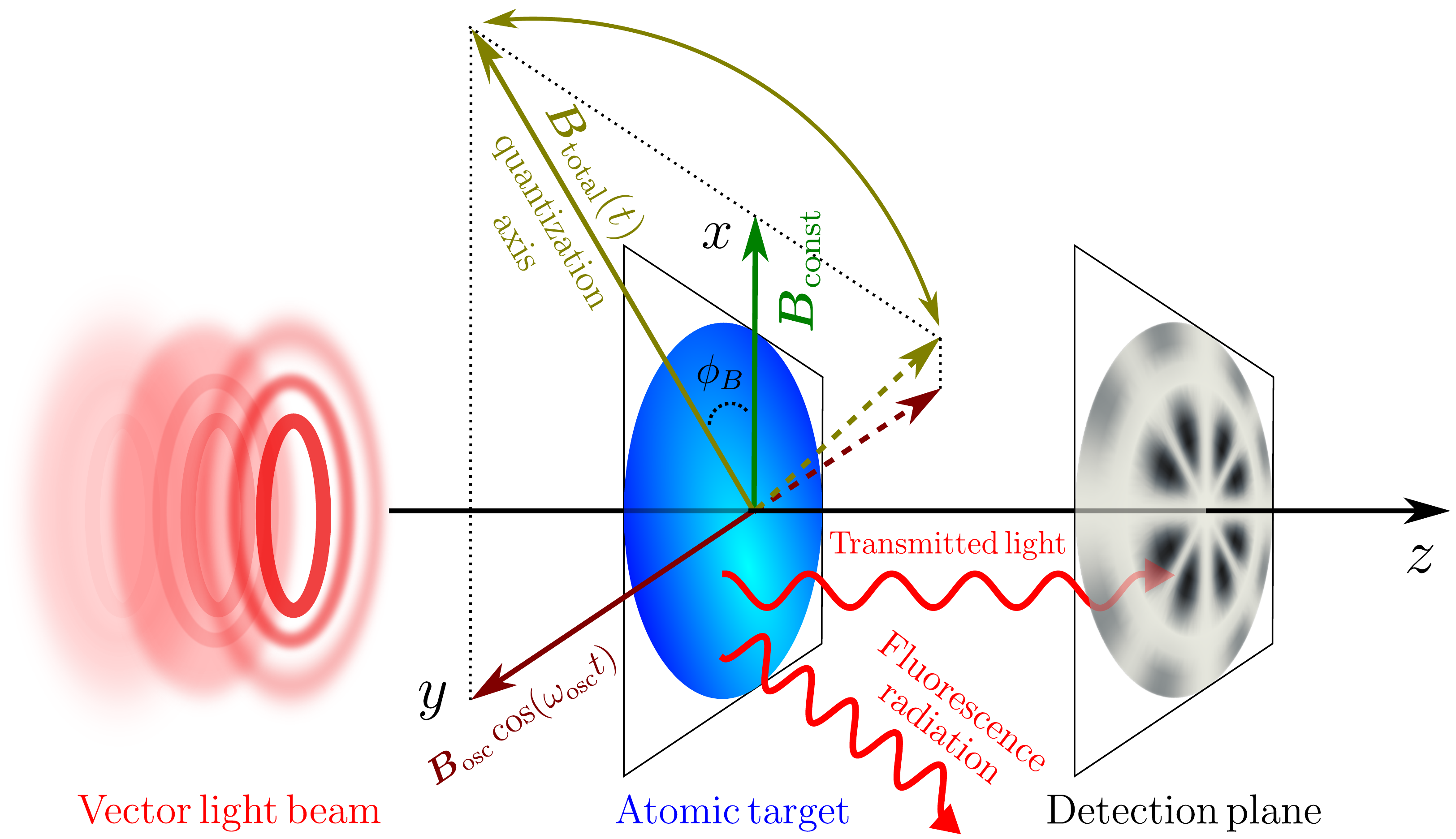}
	\caption{Proposed geometry of the experimental setup: The atomic target is subjected to the combination of a reference DC magnetic field $\bm{B}_{\mathrm{const}}$ and an AC test field $\bm{B}_{\mathrm{AC}} = \bm{B}_{\mathrm{osc}} \cos(\omega_{\mathrm{osc}} t)$, and illuminated by a vector light beam propagating along $z$. The quantization axis is chosen along the time varying direction of the total magnetic field in the $x$-$y$ plane.}
	\label{fig:geometry}
\end{figure*}

To determine the effect of the applied magnetic field on the light absorption profile, we employ density matrix theory, whose basic formulas are briefly reviewed in Sec.~\ref{Sec.Theory}. We show, in particular, how to calculate the time evolution of the atomic density matrix from the transition amplitudes for vector beams and Zeeman shifts caused by the superposition of the reference and test magnetic fields. We assume that the amplitude and polarization of the vector light change over distances much larger than an individual atom, so that we can evaluate light-matter interaction locally. Furthermore, we consider an adiabatic regime, where atomic dynamics happen at time scales much faster than changes to the light intensity (as may occur for pulsed light), and changes to the direction and intensity of the magnetic field. From the density matrix we obtain in Sec.~\ref{sec:results} the light absorption profile illustrated for the case of the $^{87}$Rb D$2$ line. We first investigate the time evolution of the absorption profiles and show that the petal-like absorption patterns rotate about the beam axis at a rate that depends on the applied AC magnetic field. In general, this effect can be used for magnetometry, but it requires a high time resolution of the detector. An easier characteristic to observe is the absorption profile averaged over the measurement time. We found that the averaged profile can be sensitive to both the strength and frequency of the test magnetic field, and this sensitivity is most pronounced for frequencies up to a hundred kHz and strengths up to several Gauss for the reference field of about one Gauss. Finally, Sec.~\ref{Sec.Summary} provides a brief summary and outlook.

\section{Theory}\label{Sec.Theory}
\subsection{Geometry of the process}\label{subsec:geometry}

\begin{figure}[t]
    \centering
    \includegraphics[width=0.45\textwidth]{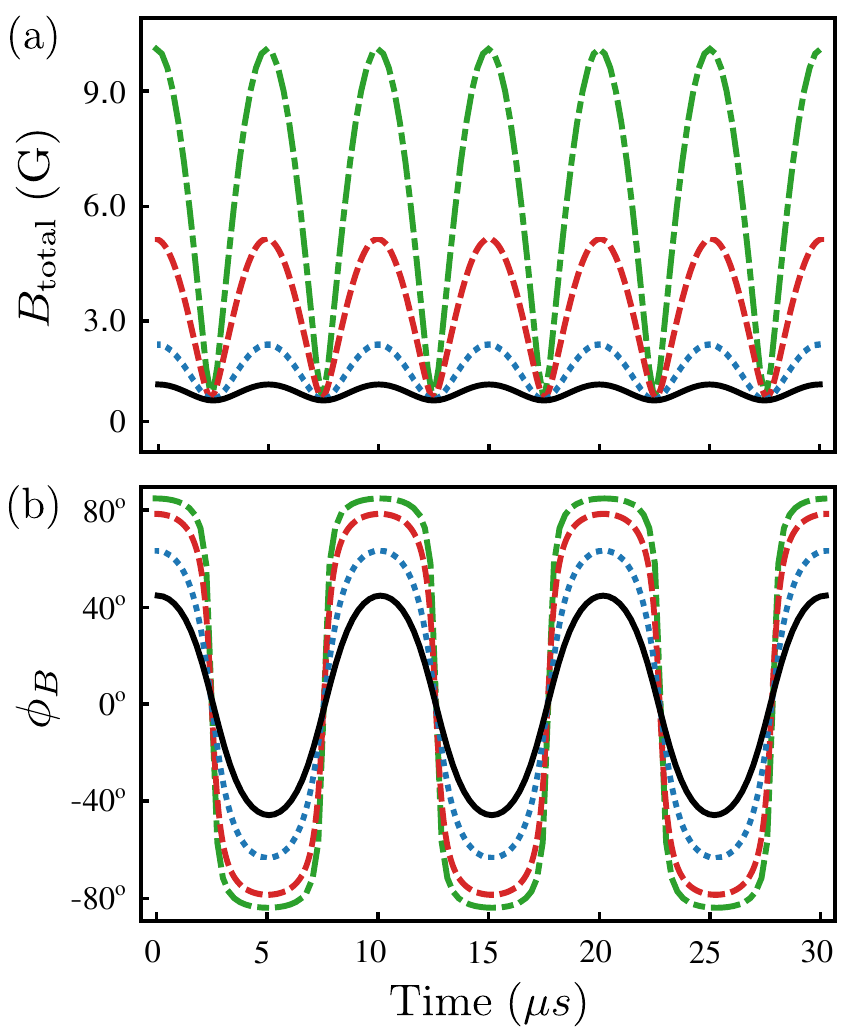}
    \caption{Time dependence of the strength (a) and orientiation (b) of the  magnetic field $\bm{B}_{\mathrm{total}} (t) = \bm{B}_{\mathrm{const}} + \bm{B}_{\mathrm{osc}} \mathrm{cos}(\omega_{\mathrm{osc}} t)$. Results are shown for $B_{\mathrm{osc}} =$ 1 G (black solid lines), $B_{\mathrm{osc}} =$ 2 G (blue dotted lines), $B_{\mathrm{osc}} =$ 5 G (red dashed lines), and $B_{\mathrm{osc}} =$ 10 G (green dash-dotted lines). For all curves, $\omega_{\mathrm{osc}} = 2\pi \times$ 100 kHz and $B_{\mathrm{const}} =$ 1 G.}
    \label{fig:twoplots}
\end{figure}

In the present work we consider the interaction of atoms with a structured light beam propagating along the $z$-axis and having the frequency $\omega$. We assume that the atoms are exposed to a combination of (reference) static and (test) oscillating magnetic fields. The static $\bm{B}_{\mathrm{const}} = B_{\mathrm{const}} \bm{e}_x$ is applied along $x$, while $\bm{B}_{\mathrm{AC}} = \bm{B}_{\mathrm{osc}}\mathrm{cos}(\omega_{\mathrm{osc}} t) =B_{\mathrm{osc}}\mathrm{cos}(\omega_{\mathrm{osc}} t) \bm{e}_y$ oscillates along the $y$-axis, as shown in Fig.~\ref{fig:geometry}. The resulting magnetic field $\bm{B}_{\mathrm{total}} (t) = \bm{B}_{\mathrm{const}} + \bm{B}_{\mathrm{osc}} \mathrm{cos}(\omega_{\mathrm{osc}} t)$ has the strength

\begin{align}
            B_{\mathrm{total}} (t) = \sqrt{B_{\mathrm{const}}^{2} + B_{\mathrm{osc}}^{2} \mathrm{cos}^{2} (\omega_{\mathrm{osc}} t)} \label{eq:Btotal}
\end{align}

\noindent oscillating between $B_{\mathrm{const}}$ and $\sqrt{B_{\mathrm{const}}^{2} + B_{\mathrm{osc}}^{2}}$, while its direction is characterized by the angle

\begin{align}
            \phi_{B}(t) = \arctan (B_{\mathrm{osc}} \mathrm{cos}(\omega_{\mathrm{osc}} t)/B_{\mathrm{const}}) \label{eq:PhiB}
\end{align}

 \noindent varying in the range $- \mathrm{arctan} (B_{\mathrm{osc}}/B_{\mathrm{const}})  \leq \phi_{B} \leq + \mathrm{arctan} (B_{\mathrm{osc}}/B_{\mathrm{const}})$. A typical variation of $B_{\mathrm{total}} (t)$ and $\phi_{B}(t)$ with $t$ is shown in Fig.~\ref{fig:twoplots}. We take fairly close values of $B_{\mathrm{const}}$ and $B_{\mathrm{osc}}$ so that the direction $\phi_{B}(t)$ of the resulting magnetic field oscillates with a sufficiently large amplitude.

By design, the field $\bm{B}_{\mathrm{total}}(t)$ is always perpendicular to the light propagation direction and oscillates around $\bm{B}_{\mathrm{const}}$. For such a complex geometry, particular attention should be paid to the choice of the quantization axis of the entire system. In our work we take the quantization axis to be along the resulting magnetic field $\bm{B}_{\mathrm{total}}(t)$, because this choice simplifies the calculation of the Zeeman splitting \cite{Bransden:2003}. On the other hand, the description of the coupling between atoms and photons becomes somewhat more complicated because the direction of $\bm{B}_{\mathrm{total}}(t)$, and hence of the quantization axis, is not stationary in the reference frame of the light beam. Of course, the observables should not depend on the specific choice of the quantization axis. To see this, we additionally performed calculations with the quantization axis along the static magnetic field $\bm{B}_{\mathrm{const}}$. Both choices of the quantization axis lead to the same result.

\subsection{Vector light beams}\label{subsec:vectortwist}
In optical physics, light beams are usually described in terms of the electric field. Such choice allows for an easier treatment of light-matter interaction in the dipole approximation. In this work we assume that the electric field of the incident vector light beam in the paraxial regime has the form:

\begin{align}
        \nonumber \mathbf{E}^{\mathrm{(vec)}}(\mathbf{r},t) =& \, E_{0} \left[\mathrm{cos}\left(2\phi\right) \bm{e}_x + \mathrm{sin}\left(2\phi\right) \bm{e}_y \right] \\
        &\times J_{2}(\varkappa r_{\bot}) e^{ik_{z}z} e^{-i\omega t} , \label{eq:EFieldparaxial}
\end{align}

\noindent where $E_0$ is a constant amplitude, $\omega$ is its frequency, $r_\perp$, $\phi$, and $z$ are cylindrical coordinates, $k_z$ and $\varkappa$ are longitudinal and transverse components of the linear momentum, respectively, and $J_{n}(\varkappa r_\perp)$ is the Bessel function. In order to develop a general and relativistic theory as well as to simplify the discussion, it is more convenient to describe the incident radiation in terms of the vector potential. For example, in order to obtain electric field~\eqref{eq:EFieldparaxial}, one has to start from the vector potential

\begin{align}
    \bm{A}^{\mathrm{(vec)}}(\bm{r},t) =  \frac{1}{\sqrt{2}} &\left[ \bm{A}^{\mathrm{(B)}}_{-m_{\gamma}, \, \lambda = +1}(\bm{r},t) - \bm{A}^{\mathrm{(B)}}_{m_{\gamma}, \, \lambda = -1}(\bm{r},t)  \right] \, \label{eq:vector}
\end{align}

\noindent which is a linear combination of two Bessel beams. Since these wave solutions with an annular intensity structure have been frequently discussed in the past~\cite{Knyazev/PU:2018, Matula/JPB:2013, Schulz/PRA:2020}, we may restrict ourselves to a rather short account of basic formulas. The Bessel beam is characterized by the well-defined helicity $\lambda$ and the projection $m_{\gamma}$ of the total angular momentum upon the propagation direction. Moreover, its longitudinal momentum $k_{z}$ and absolute value of the transverse momentum $\varkappa = |\bm{k}_{\bot}|$ are also fixed, see Ref.~\cite{Matula/JPB:2013} for further details. The vector potential for the Bessel beam can be written as

\begin{align}
\bm{A}^{\mathrm{(B)}}_{m_{\gamma},\lambda}(\bm{r}) = A_{0}\int a_{\varkappa m_{\gamma}}(\bm{k}_{\bot}) \;\bm{e}_{\bm{k}\lambda} e^{i\bm{k}\cdot\bm{r}} e^{-i\omega t} \; \frac{d^{2}\bm{k}_{\bot}}{(2\pi)^2},\label{eq:mode}
\end{align}

\noindent where $A_{0}=E_0 / \omega$ and $a_{\varkappa m_{\gamma}}(\bm{k}_{\bot})$ is a weight function given by:

\begin{align}
    a_{\varkappa m_{\gamma}}(\bm{k}_{\bot}) = \frac{2\pi}{\varkappa} (-i)^{m_{\gamma}} e^{im_{\gamma}\phi_{k}} \delta(k_{\bot}-\varkappa).
    \label{eq:bessel}
\end{align}
 
\noindent It follows from these expressions that the Bessel beam can be seen as a superposition of plane waves whose wave vectors $\bm{k} = (\bm{k}_{\bot},k_{z})$ are uniformly distributed upon the surface of a cone with a polar opening angle $\theta_{k} = \mathrm{arctan}(\varkappa/k_{z})$.

\begin{figure}[t]
    \centering
    \includegraphics[width=0.48\textwidth]{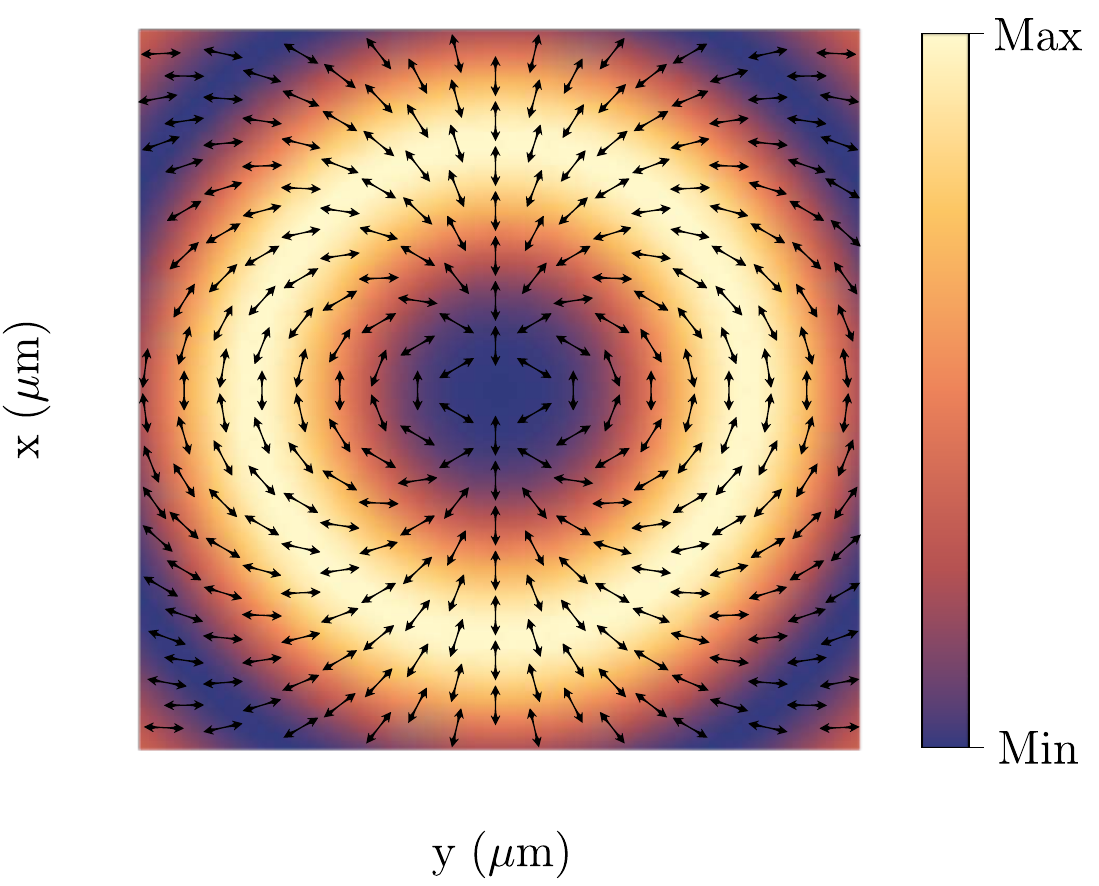}
    \caption{The intensity profile (displayed as a density plot) and polarization pattern (black arrows) of the vector beam \eqref{eq:paraxial} in the paraxial approximation with $\theta_{k} = 0.11\degree$ and $\omega = 2\pi \times 384$ THz.}
    \label{fig:intensity}
\end{figure}

\par The light field \eqref{eq:vector} is often called the vector beam because its polarization pattern is varying across the profile \cite{Rubinsztein-Dunlop/JO:2017, Gbur:2017}. In this work we consider the vector beam consisting of Bessel modes \eqref{eq:mode} with the total angular momentum projections $m_{\gamma} = \pm 1$. For arbitrary values of the opening angle $\theta_k$, the vector beam exhibits spatially dependent polarization along all three axes. Atomic physics experiments, however, commonly use vector beams produced in the paraxial regime where the transverse momentum of the photon is much smaller than the longitudinal momentum, $\varkappa \ll k_{z}$ \cite{Matula/JPB:2013}. In this regime $\theta_k$ is small and the vector potential \eqref{eq:vector} can be considerably simplified to:

\begin{align}
        \nonumber \mathbf{A}^{\mathrm{(vec)}}(\mathbf{r},t) \approx& A_{0} (-i) \left[\mathrm{cos}\left(2\phi\right) \bm{e}_x + \mathrm{sin}\left(2\phi\right) \bm{e}_y \right] \\
        &\times J_{2}(\varkappa r_{\bot}) e^{ik_{z}z} e^{-i\omega t} . \label{eq:paraxial}
\end{align}

\noindent Let us remark here that the vector beam in the paraxial approximation~\eqref{eq:paraxial} has azimuthally varying linear polarization that lies in the transverse $x$-$y$ plane \cite{Qiu/PR:2021}. The corresponding spatial field distribution is depicted in Fig.~\ref{fig:intensity}. From the vector potential~\eqref{eq:paraxial} and by using the standard relation $\boldsymbol{E}=-\partial_t \boldsymbol{A}$, we can finally obtain the electric field~\eqref{eq:EFieldparaxial}.

\subsection{Transition amplitudes}\label{subsec:transition}

Having discussed the vector potential for the incident light, we are ready now to examine its interaction with atoms. In particular, we will question the laser-induced transition between ground $\left| \alpha_g F_g M_g \right>$ and excited $\left| \alpha_e F_e M_e \right>$ atomic states whose properties can be traced back to the first-order matrix element

\begin{align}
   \nonumber & \, V^{\mathrm{(vec)}}_{eg} \\
   =& \, ec \; \left\langle \alpha_{e}F_{e}M_{e} \left \vert \sum_{q} \bm{\alpha}_{q} \cdot \bm{A}^{\mathrm{(vec)}}(\bm{r}_{q}-\bm{b})\right \vert \alpha_{g} F_{g} M_{g} \right \rangle \label{eq:amp}
\end{align}

\noindent with $\boldsymbol{F} = \boldsymbol{I} + \boldsymbol{J}$, where $\boldsymbol{I}$ and $\boldsymbol{J}$ are the nuclear and electron angular momenta, respectively, $M$ is the projection of $\boldsymbol{F}$ on the quantization axis, $\alpha$ denotes all additional quantum numbers required to specify the state uniquely, $e$ is the elementary charge, and $c$ is the speed of light. Moreover, $q$ runs over all electrons in a target atom and $\bm{\alpha}_q$ denotes the vector of Dirac matrices for the $q$th particle \cite{Johnson:2007}.  In Eq.~\eqref{eq:amp}, we have introduced the impact parameter $\bm{b} = (b \cos \phi_b, b \sin \phi_b, 0)$ to specify the position of the atom within the wave front \cite{Knyazev/PU:2018}. In particular, $b=0$ corresponds to an atom located on the vortex line of a light beam. The impact parameter plays an important role in the analysis due to the complex spatial structure of the vector beam, see Fig.~\ref{fig:intensity}.

Similar to the vector potential \eqref{eq:vector}, the transition amplitude $V^{\mathrm{(vec)}}_{eg}$ can be expressed in terms of its Bessel counterparts as

\begin{align}
    \nonumber & \,V_{eg}^{\mathrm{(vec)}} \\
    =& \, \frac{1}{\sqrt{2}} \, \left[V_{eg}^{(\mathrm{B})}(-m_{\gamma},\lambda = +1) - V^{(\mathrm{B})}_{eg}(m_{\gamma}, \lambda = -1) \right]. \label{eq:bamp}
\end{align}

\noindent The evaluation of the transition amplitude for Bessel beams \eqref{eq:vector} has already been discussed in detail in Refs.~\cite{Schulz/PRA:2020, Solyanik-Gorgone/JOSAB:2019, Peshkov/AdP:2023}. For the geometry shown in Fig.~\ref{fig:geometry}, the final form of this amplitude is

\begin{align}
    &V_{eg}^{(\mathrm{B})}(m_{\gamma},\lambda) = A_{0} ec \sqrt{2\pi} \; \sum_{M} i^{L+M}  \; [L,F_{g}]^{1/2} \; (i\lambda)^{p} \, \notag \\
    &\;\;\;\;\; \times (-1)^{m_\gamma}\; e^{i(m_\gamma-M)\phi_{b}} \; J_{m_{\gamma}-M}(\varkappa b) \, d^{L}_{M,\lambda}(\theta_{k}) \, \notag \\
    &\;\;\;\;\; \times D^{L}_{M_{e} - M_{g},M}(\pi,\pi/2,\pi-\phi_{B}(t)) \, \notag \\ 
    &\;\;\;\;\; \times \langle F_{g} M_{g} L M_{e} - M_{g}| F_{e} M_{e}\rangle  \;(-1)^{I+F_{g}+L+J_{e}} \, \notag \\
    &\;\;\;\;\; \times
    \begin{Bmatrix}
        F_e & F_g & L \\
        J_g & J_e & I 
    \end{Bmatrix}
    \langle \alpha_{e}J_{e}|| 
    H_\gamma (pL) ||\alpha_{g} J_{g}\rangle,
\label{eq:TME}
\end{align}

\noindent where we have used the notation $\langle \alpha_{e}J_{e}|| H_\gamma (pL) ||\alpha_{g} J_{g}\rangle$ to denote the reduced matrix element for magnetic ($p = 0$) and electric ($p = 1$) transitions. Furthermore, the Euler angles as the arguments of the Wigner $D$-functions $d^{L}_{M,\lambda}(\theta_{k})$ and $D^{L}_{M_{e} - M_{g},M}(\pi,\pi/2,\pi-\phi_{B}(t))$ in Eq.~\eqref{eq:TME} characterize the rotation from the atomic frame with the quantization axis along the magnetic field $\bm{B}_{\mathrm{total}} (t)$ to the photon frame with the quantization axis along the wave vector $\bm{k}$ \cite{Rose:1957}. The time-dependent angle $\pi - \phi_{B}(t)$ refers to the oscillation of the polarization vector of the incident light at the position of the atom in its (time-dependent) reference frame. The transition amplitude~\eqref{eq:TME} is valid for non-paraxial light beams as it was derived from the vector potential in the form~\eqref{eq:vector} and~\eqref{eq:mode}. However, in order to explain the results of the present work we might restrict ourselves to the simpler case of paraxial light fields.

\subsection{Density-matrix formalism}\label{subsec:densitymatrix}

Due to the presence of both the incident radiation and the time-dependent magnetic field, the populations of atomic ground and excited states can vary with time. To investigate time dependence of atomic level populations, it is convenient to use the time-dependent density matrix theory~\cite{Blum:2012}. In this approach, the state of a system is represented by the density operator $\hat{\rho}(t)$ satisfying the Liouville-von Neumann equation:

\begin{align}
    \frac{d}{dt}\hat{\rho}(t) = -\frac{i}{\hbar}[\hat{H}(t),\hat{\rho}(t)] + \hat{R}(t).\label{eq:liouville}
\end{align}

\noindent Here $\hat{H}(t)$ is the total Hamiltonian of the atom in the presence of external fields, and $\hat{R}(t)$ is introduced to take into account phenomenologically spontaneous decay~\cite{Auzinsh:2010}. We consider transitions between the Zeeman sublevels of the ground $|\alpha_{g} F_{g} M_{g}\rangle$ to those of the excited state $|\alpha_{e} F_{e} M_{e}\rangle$, described by a density matrix of size $(2F_g+2F_e+2) \times (2F_g+2F_e+2)$. In this basis, we can write the elements of the density matrix as:

\begin{subequations}
    \begin{align}
    \rho_{gg^{\prime}}(t) =& \langle \alpha_{g} F_{g} M_{g}| \hat{\rho}(t)| \alpha_{g} F_{g} M^{\prime}_{g}\rangle, \\ 
    \rho_{ee^{\prime}}(t) =& \langle \alpha_{e} F_{e} M_{e}| \hat{\rho}(t)| \alpha_{e} F_{e} M^{\prime}_{e}\rangle,  \\
    \rho_{ge}(t) =& \langle \alpha_{g} F_{g} M_{g}| \hat{\rho}(t)| \alpha_{e} F_{e}M_{e}\rangle, \\
    \rho_{eg}(t) =& \langle \alpha_{e} F_{e} M_{e}| \hat{\rho}(t)| \alpha_{g} F_{g}M_{g}\rangle.
    \end{align}\label{eq:elements}
\end{subequations}

\noindent In Eqs.~\eqref{eq:elements}, the diagonal elements $\rho_{gg}(t)$ and $\rho_{ee}(t)$ are the probabilities of finding an atom in the Zeeman substates $|\alpha_{g} F_{g} M_{g}\rangle$ and $|\alpha_{e} F_{e} M_{e}\rangle$, whereas the off-diagonal elements describe the coherence between them.

In its matrix form the Liouville-von Neumann equation~\eqref{eq:liouville} represents a system of coupled differential equations for the evolution of the density matrix elements $\rho_{gg^{\prime}}(t)$, $\rho_{ee^{\prime}}(t)$, $\rho_{ge}(t)$, and $\rho_{eg}(t)$. To solve these equations, we introduce $\tilde{\rho}_{gg^{\prime}}(t) = \rho_{gg^{\prime}}(t)$, $\tilde{\rho}_{ee^{\prime}}(t) = \rho_{ee^{\prime}}(t)$, $\tilde{\rho}_{ge}(t) = \rho_{ge}(t) e^{-i\omega t}$, $\tilde{\rho}_{eg}(t) = \rho_{eg}(t) e^{i\omega t}$ and employ the rotating-wave approximation, which is valid when $\omega$ is sufficiently close to resonance~\cite{Auzinsh:2010, Wense:2020}. This approximation allows us to eliminate the fast-oscillating terms proportional to $e^{\pm 2i\omega t}$, so that we can rewrite the Liouville-von Neumann equation as

\begin{widetext}
 \begin{subequations}
 \allowdisplaybreaks
    \begin{align}
    \allowdisplaybreaks
    \frac{d}{dt}\tilde{ \rho}_{g g^\prime}(t) =& - i \Omega_g^{\mathrm{(L)}} (t) \left[M_{g} - M^{\prime}_{g} \right] \tilde{\rho}_{g g^\prime}(t) \; - \frac{i}{2\hbar} \left[ \sum_{M_e} V_{e g}^{*} \, \tilde{\rho}_{e g^\prime}(t) - \sum_{M_e} V_{e g^\prime} \, \tilde{\rho}_{g e}(t) \right] + R_{g g^\prime}(t) ,\\
    \frac{d}{dt}\tilde{\rho}_{e e^\prime}(t) = &\, - i \Omega_e^{\mathrm{(L)}} (t) \left[M_{e} - M^{\prime}_{e} \right] \tilde{\rho}_{e e^\prime}(t) \; - \frac{i}{2\hbar} \left[ \sum_{M_g} V_{e g} \, \tilde{\rho}_{g e^\prime}(t) - \sum_{M_g} V_{e^\prime g}^{*} \, \tilde{\rho}_{e g}(t) \right] + R_{e e^\prime}(t) ,\\
    \frac{d}{dt} \tilde{\rho}_{g e}(t) =& \; -i \Delta \tilde{\rho}_{g e}(t) + i \left[ \Omega_e^{\mathrm{(L)}} (t) M_{e} - \Omega_g^{\mathrm{(L)}} (t) M_{g} \right] \tilde{\rho}_{g e}(t) \; - \frac{i}{2\hbar} \left[ \sum_{M_e^\prime} V_{e^\prime g}^{*} \, \tilde{\rho}_{e^\prime e}(t) - \sum_{M_g^\prime} V_{e g^\prime}^{*} \, \tilde{\rho}_{g g^\prime}(t) \right] + R_{g e}(t) ,\\
   \frac{d}{dt} \tilde{\rho}_{eg}(t)  =& \; i \Delta \tilde{\rho}_{e g}(t) - i \left[ \Omega_e^{\mathrm{(L)}} (t) M_{e} - \Omega_g^{\mathrm{(L)}} (t) M_{g} \right] \tilde{\rho}_{e g}(t) \; - \frac{i}{2\hbar} \left[ \sum_{M_g^\prime} V_{e g^\prime} \; \tilde{\rho}_{g^\prime g}(t) - \sum_{M_e^\prime} V_{e^\prime g} \; \tilde{\rho}_{e e^\prime}(t) \right] + R_{e g}(t),
    \end{align}\label{eq:diff-eqn}
 \end{subequations}
\end{widetext}

\noindent where $\Delta = \omega - \omega_{0}$ denotes the light frequency detuning from resonance, $\omega_{0}$ is the atomic transition frequency in the absence of external fields, $\Omega^{\mathrm{(L)}} (t) = g_{F} \mu_{B} B_{\mathrm{total}} (t) /\hbar$ is the Larmor frequency, and $V_{eg}$ is the transition matrix element, which is proportional to the Rabi frequency. The contribution of spontaneous decay to the $R(t)$ terms, obtained from the rate for emission summed over polarizations and integrated over angles, is

\begin{subequations}
\allowdisplaybreaks
\begin{align}
\allowdisplaybreaks
    \nonumber &R_{g g^\prime} (t) = \;\Gamma [F_{g},J_{e}] \begin{Bmatrix}
F_e & F_g & L \\
J_g & J_e & I
\end{Bmatrix}^2 \\
    &\times \sum\limits_{M_{e}, M^{\prime}_{e}, M} \langle F_g M_{g} L M | F_e M_{e} \rangle \, \tilde{\rho}_{e e^\prime} (t) \, \langle F_g M^\prime_{g} L M | F_e M^\prime_{e} \rangle  \, , \\
    &R_{e e^\prime} (t) = \, - \Gamma [F_{g},J_{e}] \begin{Bmatrix}
F_e & F_g & L \\
J_g & J_e & I
\end{Bmatrix}^2 \tilde{\rho}_{e e^\prime} (t), \\
    &R_{g e} (t) = \, - \frac{1}{2} \Gamma [F_{g},J_{e}] \begin{Bmatrix}
F_e & F_g & L \\
J_g & J_e & I
\end{Bmatrix}^2 \tilde{\rho}_{g e} (t) \, , \\
    &R_{e g} (t) = \, - \frac{1}{2} \Gamma [F_{g},J_{e}] \begin{Bmatrix}
F_e & F_g & L \\
J_g & J_e & I
\end{Bmatrix}^2 \tilde{\rho}_{e g} (t),  
\end{align}
\label{eq:RTerms}
\end{subequations}

\noindent where $[F_{g},J_{e}] = (2F_{g}+1)(2J_{e}+1)$ and $\Gamma$ is the decay rate of the upper level $|\alpha_{e} J_{e}\rangle$ \cite{Tremblay/PRA:1990, Schmidt/PRA:2024}. Moreover, $L=1$ corresponds to a dipole transition, $L=2$ to a quadrupole, $L=3$ to an octupole, etc.

\subsection{Light absorption profile}\label{subsec:absorptionprofile}
Solving the Liouville-von Neumann equation~(\ref{eq:diff-eqn}) numerically allows the determination of the atomic density matrix at any instant of time. The elements of this matrix are directly related to physical observables. In optical magnetometry experiments with vector beams and atoms, the absorption profile of the light is most commonly observed~\cite{Castellucci/PRL:2021, Qiu/PR:2021}. Different approaches can be used to analyze such profiles. In Ref.~\cite{Castellucci/PRL:2021}, for example, the approach based on Fermi's golden rule and spatially dependent partially dressed states has been successfully applied to explain the experimental findings. We take a different approach here in which we focus on the diagonal density matrix elements $\rho_{ee}(t)$ representing the population of photoexcited atomic states \cite{Blum:2012}. The method relies on a simple assumption that atoms excited to the upper state must decay back to the ground state by the emission of photons in all directions. As a result, regions with many excited atoms appear darker than those with less excitations as the detector measures the intensity of light in the direction of the incoming beam, see Fig.~\ref{fig:geometry}. In other words, high values of $\rho_{ee}$ imply a large imaginary part of the refractive index of the medium. Thus the analysis of the light absorption profile may be reduced to the analysis of the density matrix elements $\rho_{ee}$ which depend on the position $\bm{b}$ of the target atom through the transition amplitudes $V^{\mathrm{(vec)}}_{eg}$, as well as on the properties of the magnetic field through the Larmor frequencies $\Omega^{\mathrm{(L)}}$ and the angle $\phi_{B}$.


\section{Results and Discussion}\label{sec:results}

\begin{figure}[t]
    \centering
    \includegraphics[width=0.48\textwidth]{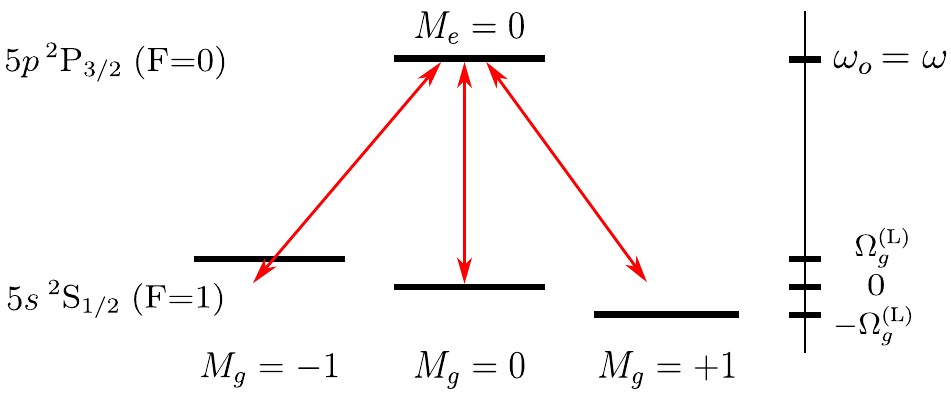}
    \caption{The $5s \;\, {}^2 S_{1/2}$ ($F=1$) $-$ $5p \;\, {}^2 P_{3/2}$ ($F=0$) transition in $^{87}$Rb. The lower sublevels are split by the energy $\hbar \Omega_g^{\mathrm{(L)}}$ as given by the Larmor frequency of the atom in the magnetic field. The arrows indicate the interaction with light at zero detuning.}
    \label{fig:energy_levels}
\end{figure}

\begin{figure*}
    \centering
    \includegraphics[width=0.99\textwidth]{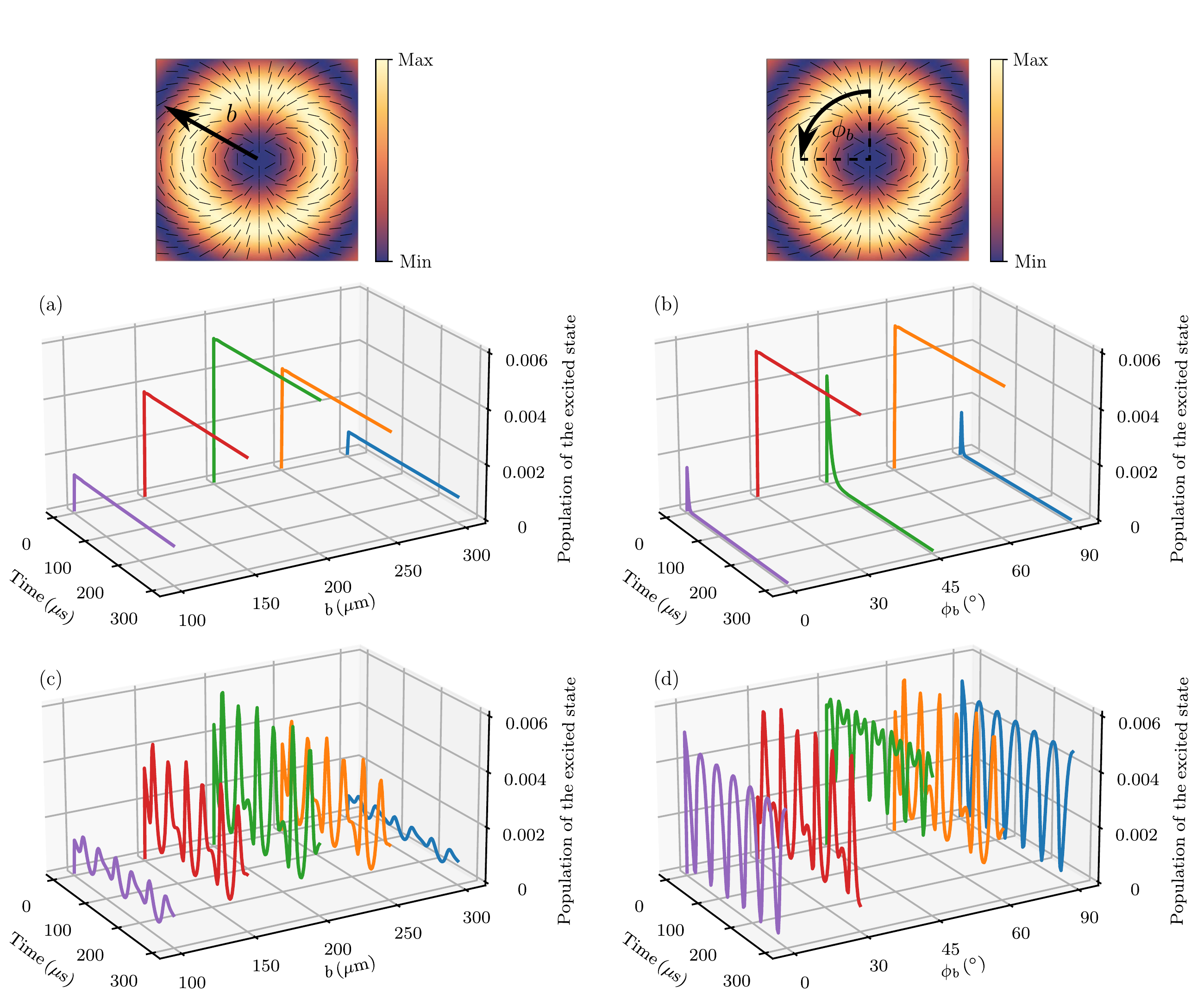}
    \caption{Population \eqref{eq:population} of the $5p \;\, {}^2 P_{3/2}$ ($F=0$) excited state of ${}^{87}$Rb as a function of time and impact parameter $\boldsymbol{b}$. The atom interacts with the vector beam \eqref{eq:vector} in the presence of either the static magnetic field with $B_\mathrm{const}=1$ G and $B_\mathrm{osc}=0$ (a),(b) or the AC magnetic field with $B_\mathrm{const}=B_\mathrm{osc}=1$ G and $\omega_\mathrm{osc}=2\pi \times 10$ kHz (c),(d). Calculations were performed for different positions $\bm{b}$ of the atom relative to the beam center (see the top row of the figure). (a),(c) Dependence on $b$ for $\phi_b = 60\degree$; (b), (d) Dependence on $\phi_b$ for $b=200$ $\mu$m. For all curves, $\Delta =0$, $m_{\gamma} = \pm 1$, $\theta_k = 0.11\degree$, and $A_0 = 3.07 \times 10^{-14}$.}
    \label{fig:impact}
\end{figure*}

In the previous section we have outlined the necessary theory for describing the interaction of vector beams with atoms in the presence of an external magnetic field. This formalism can be applied to analyze transitions between two arbitrary hyperfine-structure levels over a wide frequency range. Here we focus on the $5s \;\, {}^2 S_{1/2}$ ($F=1$) $-$ $5p \;\, {}^2 P_{3/2}$ ($F=0$) electric dipole (E1) transition in $^{87}$Rb at zero detuning (i.e.~$\omega = \omega_{0} = 2 \pi \times 384$ THz, see Fig.~\ref{fig:energy_levels}). This transition has already been utilized in the atomic magnetometer based on vector beams \cite{Castellucci/PRL:2021}. The atom is assumed to be initially unpolarized. The required reduced matrix element $\left< 5p \;\, {}^2 P_{3/2} || H_\gamma (E1) || 5s \;\, {}^2 S_{1/2} \right>$ and the spontaneous decay rate $\Gamma$ have been calculated using the JAC code \cite{Fritzsche/CPC:2019}. Here the theoretically obtained value $\Gamma_\mathrm{theo}=4.042 \cdot 10^7$ s${}^{-1}$ is relatively close to the measured decay rate $\Gamma_\mathrm{exp}=3.811 \cdot 10^7$ s${}^{-1}$ of the ${}^2 P_{3/2}$ excited state \cite{Volz/PHysScripta:1996}. In what follows we shall only deal with the vector potential \eqref{eq:vector} with the total angular momentum projections $m_{\gamma} = \pm 1$. We have chosen the parameters $A_0 = 3.07 \times 10^{-14}$ and $\theta_k = 0.11\degree$ such that the Bessel solution \eqref{eq:vector} reproduces the experimentally realistic Laguerre-Gaussian mode of waist $200$ $\mu$m and total power $0.4$ $\mu$W in the vicinity of the beam center. Such choice of parameters produce Rabi frequencies in the range of MHz in the regions of high beam intensity. We assume that the magnetic field frequency $\omega_\mathrm{osc}$ is much smaller than these Rabi frequencies and the decay rate so that the atom-light interaction can be considered adiabatic. Moreover, we also suppose that $\omega_\mathrm{osc}$ is much smaller than the Larmor frequency $\Omega^{\mathrm{(L)}} (t)$, which is in the range of MHz as well. For this reason we neglect magnetic field induced transitions, that can occur between Zeeman substates within the same level.

\begin{figure*}[t]
    \centering
    \includegraphics[width=0.98\textwidth]{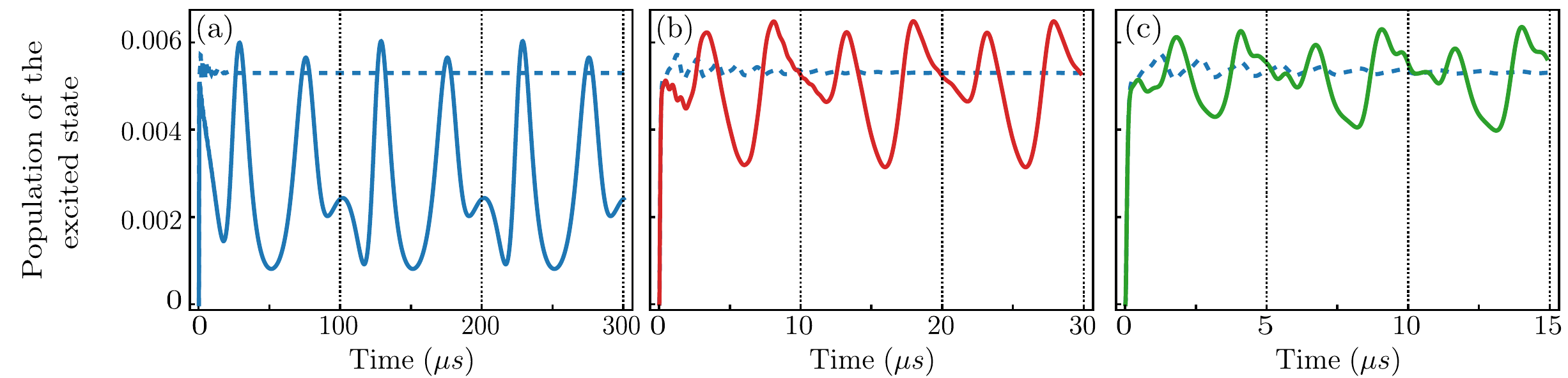}
    \caption{Excited-state populations for different frequencies of the magnetic field: (a) $\omega_{\mathrm{osc}} = 2\pi \times 10$ kHz, (b) $\omega_{\mathrm{osc}} = 2\pi \times 100$ kHz, and (c) $\omega_{\mathrm{osc}} = 2\pi \times 200$ kHz. It is assumed that $b=200$ $\mu$m, $\phi_b=60\degree$, $B_\mathrm{const}=B_\mathrm{osc}=1$ G (solid lines), $B_\mathrm{const}=1$ G and $B_\mathrm{osc}=0$ (dashed lines). Vertical bars indicate period of magnetic field oscillations $T_B$. All other parameters are the same as in Fig.~\ref{fig:impact}.}
    \label{fig:impact_freq}
\end{figure*}

\subsection{Time evolution of the excited-state population}\label{subsec:impactplot}

We start our discussion by considering the time evolution of the excited-state population

\begin{align}
 \rho_{ee}(t) = \left\langle 5p \;\, {}^2 P_{3/2} \, (F=0) \left \vert \hat{\rho}(t) \right \vert 5p \;\, {}^2 P_{3/2} \, (F=0) \right\rangle \,  
 \label{eq:population}
\end{align}

\noindent for several selected positions $\bm{b} = (b \cos \phi_b, b \sin \phi_b, 0)$ of the target atom with respect to the zero-intensity center of the vector beam. The results of the calculation are shown in Fig.~\ref{fig:impact}. In the left column, the azimuthal angle is fixed at $\phi_b = 60\degree$ and the radial distance is varied ($b=100$, $150$, $200$, $250$, and $300$ $\mu$m). In the right column, the radial distance is fixed rather at $b=200$ $\mu$m and the azimuthal angle is varied ($\phi_b = 0\degree$, $30\degree$, $45\degree$, $60\degree$, and $90\degree$). The subfigures (a) and (b) indicate the population $\rho_{ee}(t)$ of the exited state for the atoms in the external magnetic field with $B_\mathrm{const}=1$ G and $B_\mathrm{osc}=0$. For this static magnetic field, $\rho_{ee}(t)$  reaches a steady state after several tens of microseconds irrespective of the atomic position. The explicit value of the steady-state population $\rho_{ee}$ is, however, very sensitive to $b$ and $\phi_b$. Here the dependence of $\rho_{ee}$ on $b$ is mainly due to the radial distance dependence of the light intensity. For example, $\rho_{ee}$ is greater at $b=200$ $\mu$m than at $b=100$ $\mu$m and $b=300$ $\mu$m, since the light intensity is higher at this point. The variation of the excited-state population with the azimuthal angle $\phi_b$ can in turn be understood with the help of Fermi’s golden rule for the electric dipole transition rate

\begin{align}
    W \propto \left| V_{eg} \right|^2 \approx \left| \left\langle \alpha_{e}F_{e}M_{e} \left \vert \bm{e} \cdot \bm{d} \right \vert \alpha_{g} F_{g} M_{g} \right\rangle \right|^2 \, ,
    \label{eq:FermisGoldenRule}
\end{align}

\noindent where $\bm{d}$ is the dipole moment operator of the atom \cite{Bransden:2003}. As seen from Eq.~\eqref{eq:FermisGoldenRule}, the absorption depends on the direction $\bm{e}$ of local polarization of light with respect to the quantization axis given by the magnetic field. Since the polarization of the vector beam \eqref{eq:vector} varies with the azimuthal angle, the transition rate, and hence $\rho_{ee}$, is different for different $\phi_b$. For instance, $\rho_{ee}$ is much lower at $\phi_b=90\degree$ than at $\phi_b=30\degree$ or $60\degree$. A more detailed discussion of this angular dependence with explicit expressions for $W$ can be found in the work \cite{Castellucci/PRL:2021}.

\begin{figure}[b]
    \centering
    \includegraphics[width=0.47\textwidth]{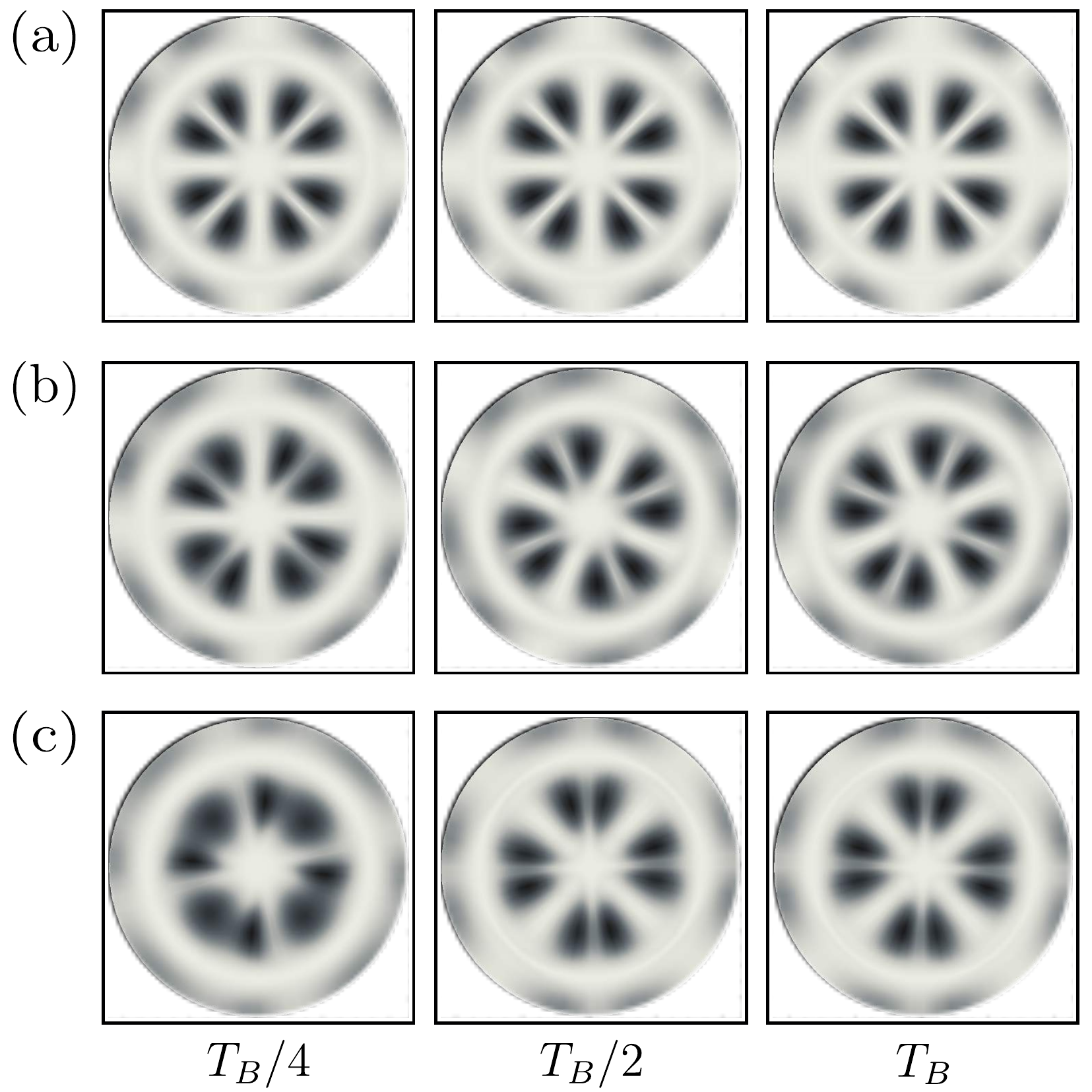}
    \caption{Absorption profile of the vector beam at different times $t = T_B/4$, $T_B/2$, and $T_B$ for rubidium atoms immersed in (a) the static magnetic field with $B_{\mathrm{const}}=1$ G, $B_{\mathrm{osc}}=0$ and the time-dependent magnetic field with (b) $B_{\mathrm{const}}=B_{\mathrm{osc}}=1$ G, $\omega_\mathrm{osc} = 2 \pi \times 10$ kHz, and (c) $B_{\mathrm{const}}=1$ G, $B_{\mathrm{osc}}=10$ G, $\omega_\mathrm{osc} = 2 \pi \times 18$ kHz. The darker shading indicates the maximum absorption and the brighter shading represents the minimum absorption. Parameters are the same as in Fig.~\ref{fig:impact}.}
    \label{fig:absoption}
\end{figure}

In contrast to the static magnetic field, the density matrix never reaches a steady state when the atoms are immersed in a time-dependent magnetic field. This is illustrated in Fig.~\ref{fig:impact}~(c)~and~(d) for $\omega_\mathrm{osc}= 2\pi \times 10$ kHz and $B_{\mathrm{const}} = B_{\mathrm{osc}} =1$ G, i.e. $-45\degree \leq \phi_{B}(t) \leq +45\degree$. Here the excited-state population $\rho_{ee}(t)$ undergoes oscillations whose character strongly depends on the atomic position. Moreover, the period of these oscillations is shorter than the period of magnetic field oscillations $T_{B} = 2\pi/\omega_\mathrm{osc} = 100$ $\mu$s. As seen from Fig.~\ref{fig:impact_freq}, the shape of the oscillations of $\rho_{ee}(t)$ changes also with the frequency of the applied magnetic field, and this is most noticeable when $\omega_\mathrm{osc}$ increases from $2\pi \times 10$ kHz to $2\pi \times 100$ kHz.

\subsection{Time evolution of the light absorption profile}\label{subsec:densityplot}

Let us next discuss the excited-state population $\rho_{ee}(t)$ in the entire $x$-$y$ plane. According to the discussion in Sec.~\ref{subsec:absorptionprofile}, this distribution can be related to the light absorption profile. Fig.~\ref{fig:absoption} shows such profiles at different times, $t = T_B/4$, $T_B/2$, and $T_B$. One should notice that $T_B$ is depending on $\omega_\mathrm{osc}$, and hence each subfigure features different moments in time. In the figure, the rate of absorption is indicated by shading: lighter shades are minimum values (weak absorption) and darker shades are maximum values (strong absorption). For the case of an external static magnetic field, we observe a stationary petal-like pattern, see Fig.~\ref{fig:absoption} (a). This result agrees well with experiment and previous theoretical predictions \cite{Castellucci/PRL:2021}. If the atoms are additionally exposed to an oscillating magnetic field, the petal-like pattern begins to rotate about the beam axis. As seen from Figs.~\ref{fig:absoption} (b) and (c), the rotation itself depends on the field parameters. This dependence can be used, for example, to determine the strength $B_\mathrm{osc}$ and frequency $\omega_{\mathrm{osc}}$ of an unknown oscillating component of the magnetic field if the static component $B_\mathrm{const}$ is known.

\subsection{Time-averaged light absorption profile}\label{subsec:timeavgplot}

\begin{figure}[t]
        \centering
        \includegraphics[width=0.45\textwidth]{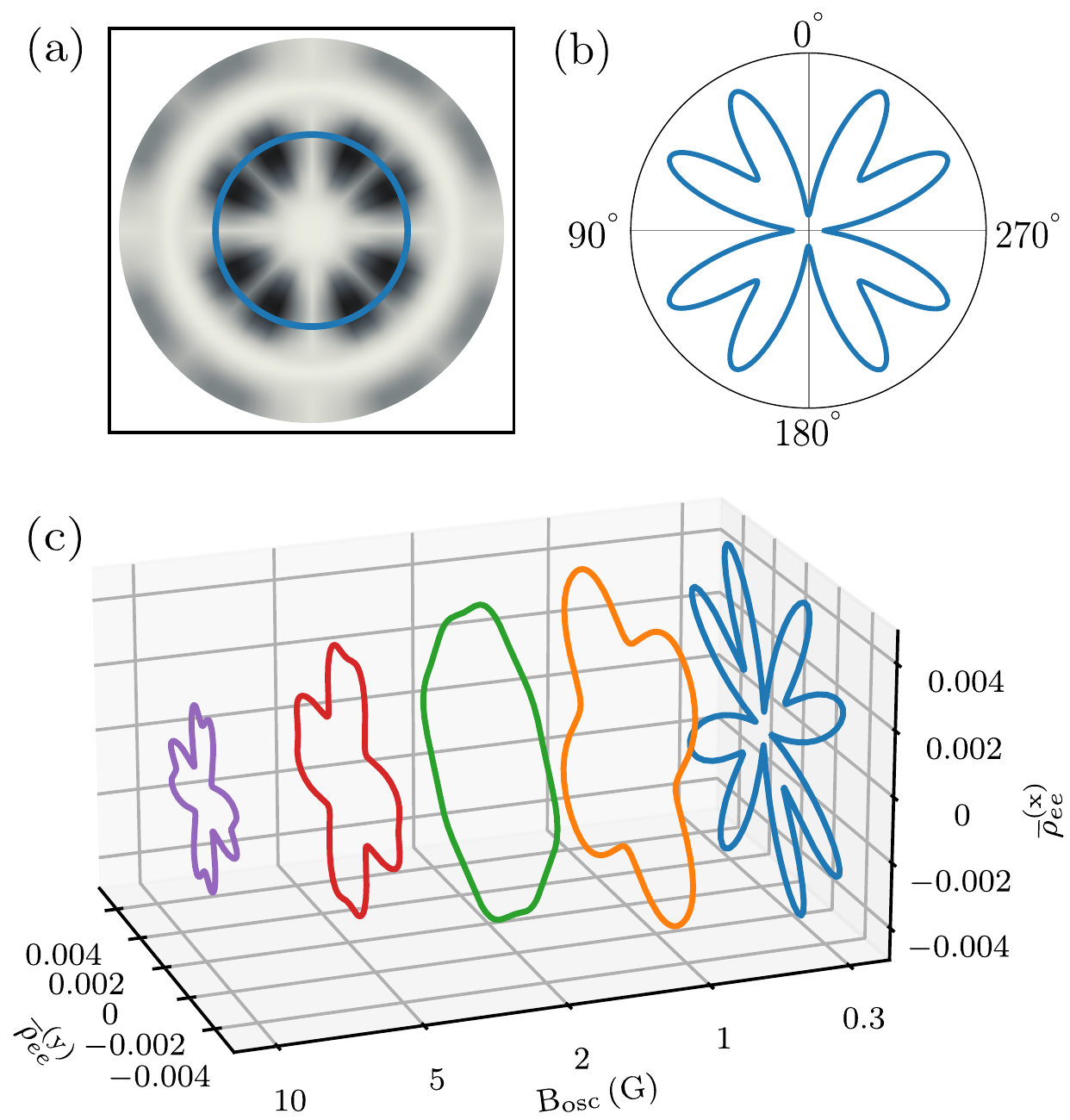}
        \caption{(a) Time-averaged absorption profile of the vector beam in the presence of the time-dependent magnetic field with $B_\mathrm{const}=1$ G, $B_\mathrm{osc}=0.3$ G, and $\omega_\mathrm{osc}=2 \pi \times 100$ kHz. All other parameters are the same as in Fig.~\ref{fig:impact}. (b) Polar plot of the absorption intensity which refers to the radius of the circle in (a) at $b=200$ $\mu$m. (c) Polar plots of absorption intensity as a function of $B_{\mathrm{osc}}$. Here $B_\mathrm{const}=1$ G and $\omega_\mathrm{osc}=2 \pi \times 100$ kHz.}
        \label{fig:timeavgDensityPlotStrength}
\end{figure}

Direct observation of the above-mentioned rotation of the petal-like absorption pattern can be difficult because of the limited time resolution of typical detectors and the requirement that the experiment would need to be executed repeatedly to obtain the time traces. The light absorption profile averaged over a cycle of oscillation

\begin{align}
    \overline{\rho}_{ee}= \frac{1}{T_B} \int_{T_B} \rho_{ee} (t) \, dt
    \label{eq:TimeAveragedPopulation}
\end{align}

\noindent can be considered a more realistic characteristic to measure. It is worth noting that since oscillations of $\rho_{ee}(t)$ repeat themselves at each cycle of magnetic field oscillations, averaging over one period $T_B$ and averaging over many periods give the same results. The time-averaged light absorption profile for $B_\mathrm{const}=1$ G, $B_\mathrm{osc}=0.3$ G, and $\omega_\mathrm{osc}=2 \pi \times 100$ kHz as well as the polar plot of absorption intensity at $b=200$ $\mu$m are displayed in Fig.~\ref{fig:timeavgDensityPlotStrength}~(a)~and~(b), respectively. Fig.~\ref{fig:timeavgDensityPlotStrength}~(c) similarly shows polar plots of $\overline{\rho}_{ee}$ at different strengths $B_\mathrm{osc}$ of the oscillating component of the magnetic field, assuming that $\omega_\mathrm{osc}=2 \pi \times 100$ kHz and $B_\mathrm{const}=1$ G. The petals adjacent to each other in the time-averaged profile begin to merge as $B_\mathrm{osc}$ varies from $0.3$ to $1$ G. With a further increase of the magnetic field strength up to $B_\mathrm{osc}=5$ G, the pattern rotates by $45\degree$. This effect can be explained by the time evolution of the magnetic field direction $\phi_B$. For $B_\mathrm{osc} > 2$ G the time evolution of $\phi_B$ resembles a rectangle function where most of the time the magnetic field is almost parallel to the $y$-axis, see Fig.~\ref{fig:twoplots}. In the case of such a magnetic field, absorption of the incident radiation~\eqref{eq:vector} is most pronounced for atoms at the angles $\phi_b=0^\circ$ and $90^\circ$, as follows from Eq.~\eqref{eq:FermisGoldenRule}. In contrast, for $B_\mathrm{osc} \leq 2$ G the direction $\phi_B(t)$ oscillates sinusoidal around the $x$-axis. Here stronger absorption of \eqref{eq:vector} occurs nearby $\phi_b=45^\circ$. This is especially visible in the polar plots in Fig.~\ref{fig:timeavgDensityPlotStrength}~(c). An even larger enhancement of the magnetic field component $B_\mathrm{osc}$, however, will not significantly affect the absorption profile $\overline{\rho}_{ee}$. Such loss of sensitivity is explained by the fact that the magnetic field direction $\phi_B(t)$ ceases to depend on $B_\mathrm{osc}$ at high $B_\mathrm{osc}$, see Fig.~\ref{fig:twoplots}.

\begin{figure}[t]
        \centering
        \includegraphics[width=0.45\textwidth]{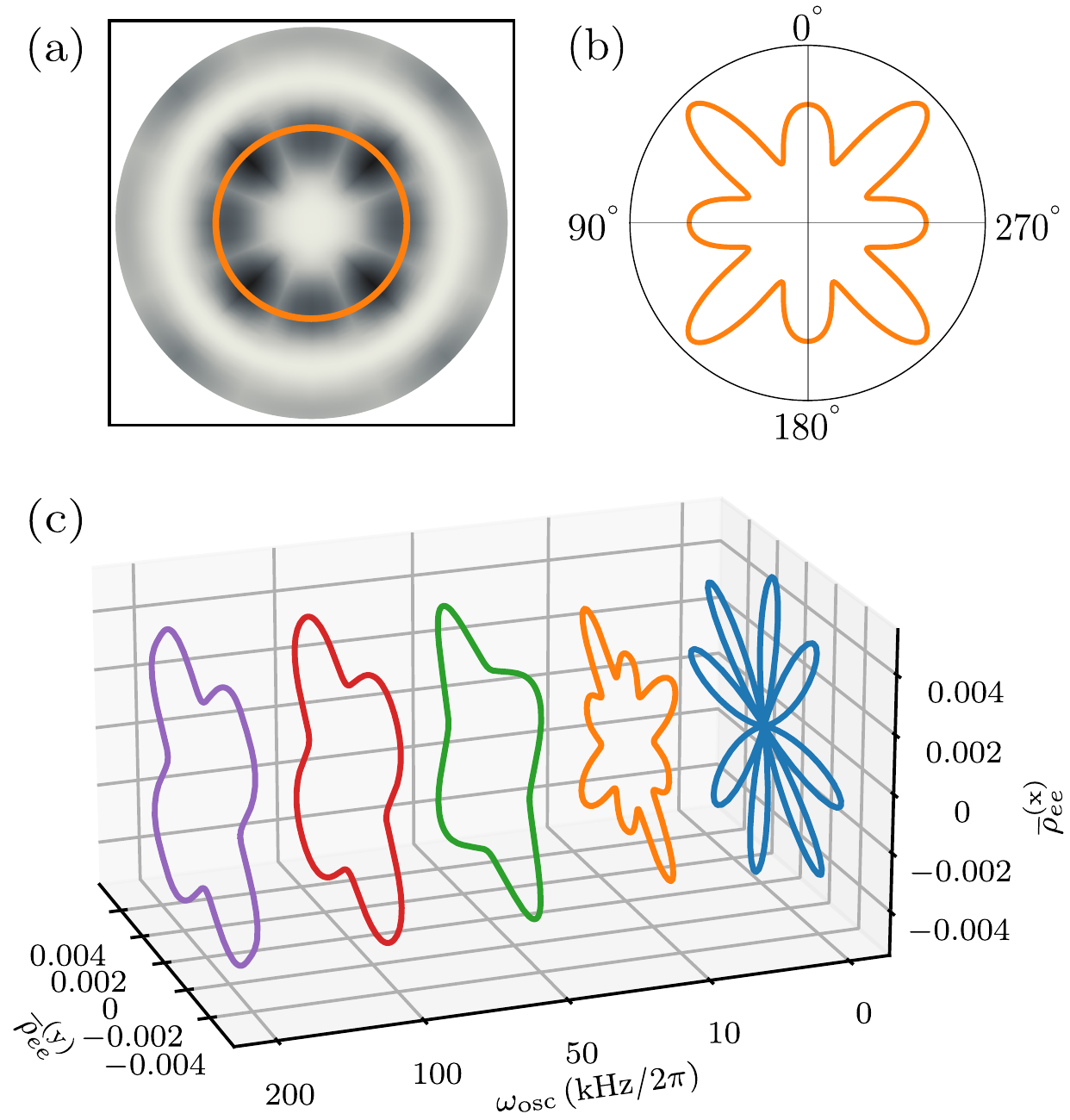}
        \caption{(a) Time-averaged absorption profile of the vector beam in the presence of the time-dependent magnetic field with $B_\mathrm{const}=B_\mathrm{osc}=1$ G and $\omega_\mathrm{osc}=2 \pi \times 10$ kHz. All other parameters are the same as in Fig.~\ref{fig:impact}. (b) Polar plot of the absorption intensity which refers to the radius of the circle in (a) at $b=200$ $\mu$m. (c) Polar plots of absorption intensity as a function of $\omega_{\mathrm{osc}}$. Here $B_\mathrm{const}=B_\mathrm{osc}=1$ G.}
        \label{fig:timeavgDensityPlotFrequency}
\end{figure}

We are also interested in how the magnetic field frequency $\omega_{\mathrm{osc}}$ affects the time-averaged absorption profile $\overline{\rho}_{ee}$. It can be seen from Fig.~\ref{fig:timeavgDensityPlotFrequency} that the minima of $\overline{\rho}_{ee}$ become less pronounced as $\omega_{\mathrm{osc}}$ increases from zero in the case of $B_\mathrm{const}=B_\mathrm{osc}=1$ G. We also see that the absorption profile ceases to change noticeably at $\omega_\mathrm{osc} > 2 \pi \times 100$ kHz. Thus the determination of the magnetic field based on the analysis of $\overline{\rho}_{ee}$ is difficult at such high magnetic field frequencies.

In the calculations above we have investigated the sensitivity of absorption profile to the strength and frequency of the oscillating magnetic field for a particular choice of computational parameters. For instance, in Figs.~\ref{fig:timeavgDensityPlotStrength}~and~\ref{fig:timeavgDensityPlotFrequency} we have chosen $B_\mathrm{const}=1$ G and $0.3$ G $\leq B_\mathrm{osc} \leq$ $10$ G. The conclusion based on these results, however, can be generalized on a wider parameter range. In particular, we performed additional calculations which indicated that sensitivity to the AC magnetic field strength and frequency is most pronounced for the case when strengths of reference and test field are comparable to each other, i.e. when $0.1 \lessapprox B_\mathrm{const} / B_\mathrm{osc} \lessapprox 10$.

In the present study, moreover, we have mainly focused on the scenario where the static (reference) and the oscillating (test) magnetic field are perpendicular to each other. As mentioned above, our theoretical approach is general and can be used for any angle between both fields. For the sake of brevity, we will not discuss here results for different field directions in detail and just mention briefly the most important findings. Namely, we found that time-averaged absorption profile is not sensitive to the strength and frequency of the oscillating magnetic field when $\bm{B}_{\mathrm{const}} \parallel \bm{B}_{\mathrm{AC}}$. This sensitivity grows with the angle between both fields and reaches its maximum when $\bm{B}_{\mathrm{const}} \perp \bm{B}_{\mathrm{AC}}$. Such dependence on the angle between both magnetic fields may allow to determine the direction and properties of the AC test field, but this would require application of the DC reference field at several directions and measurements of the absorption profile.


\section{Summary and outlook}\label{Sec.Summary}

In this paper we have studied the propagation of a vector light beam through an atomic target, exposed to an external magnetic field having static and oscillating components. Special attention was paid to the absorption profile of transmitted light, as it is often measured in experiments. To compute the absorption profile, we used the density matrix theory. The resulting expressions are general and applicable to any atomic system. As an example, calculations have been performed for the $5s \;\, {}^2S_{1/2}$ ($F=1$) $-$ $5p \;\, {}^2 P_{3/2}$ ($F=0$) transition in $^{87}$Rb. These calculations have shown that the absorption profile of a vector beam varies with time, and its temporal evolution depends on the parameters of the oscillating magnetic field. A measurable signature of the temporal changes to the population of $\rho_{ee}$ and hence the absorption can be found in the time-averaged absorption profile. We find that the averaged absorption profile is sensitive to both the strength and frequency of the oscillating magnetic field. For the static field of about one Gauss, the highest sensitivity was observed at field strengths in the range from zero to several Gauss and at frequencies in the range from zero to a hundred kHz.

Our study indicates that measurements of the absorption profile of vector light beams can be utilized to diagnose oscillating magnetic fields. The combination with a DC (reference) field enables one to extract information about frequency and strength of an AC (test) field. Based on our calculations, we found that sensitivity to both these parameters is most pronounced when the ratio $B_\mathrm{const} / B_\mathrm{osc}$ is close to unity.

In the present publication several assumptions have been made to simplify our theoretical treatment. In particular, we restricted our work to (i) the ${}^{87}$Rb D2 line induced by (ii) vector mode with particular polarization pattern given by Eq.~\eqref{eq:vector}, interacting with (iii) cold atoms whose center-of-mass motion was neglected. While these assumptions are feasible for analysis of current experiments, they have to be questioned for optimizing future measurement setups. In a forthcoming study, we therefore plan to investigate coupling of hot atomic gas with vector light mode exhibiting richer polarization structure and pay special attention to operation of hyperfine transitions.

\section*{Acknowledgments}
We acknowledge support from the Research School of Advanced Photon Science of the Helmholtz Institute Jena, HPC cluster DRACO of FSU Jena, QuantERA II Programme with funding received via the EU H2020 research and innovation programme under Grant No. 101017733, EPSRC under Grant No. EP/Z000513/1 (V-MAG), and Deutsche Forschungsgemeinschaft (DFG, German Research Foundation) under Germany's Excellence Strategy- EXC-2123 QuantumFrontiers-390837967. The authors are grateful for fruitful discussions with K.~Essink, N.~Huntemann, and E.~Peik.



\bibliography{Bibliography}

\end{document}